\documentclass{moriond}

\def\be{\begin{equation}}
\def\ee{\end{equation}}
\def\bea{\begin{eqnarray}}
\def\eea{\end{eqnarray}}

\newcommand{\Photo}{\includegraphics[height=35mm]{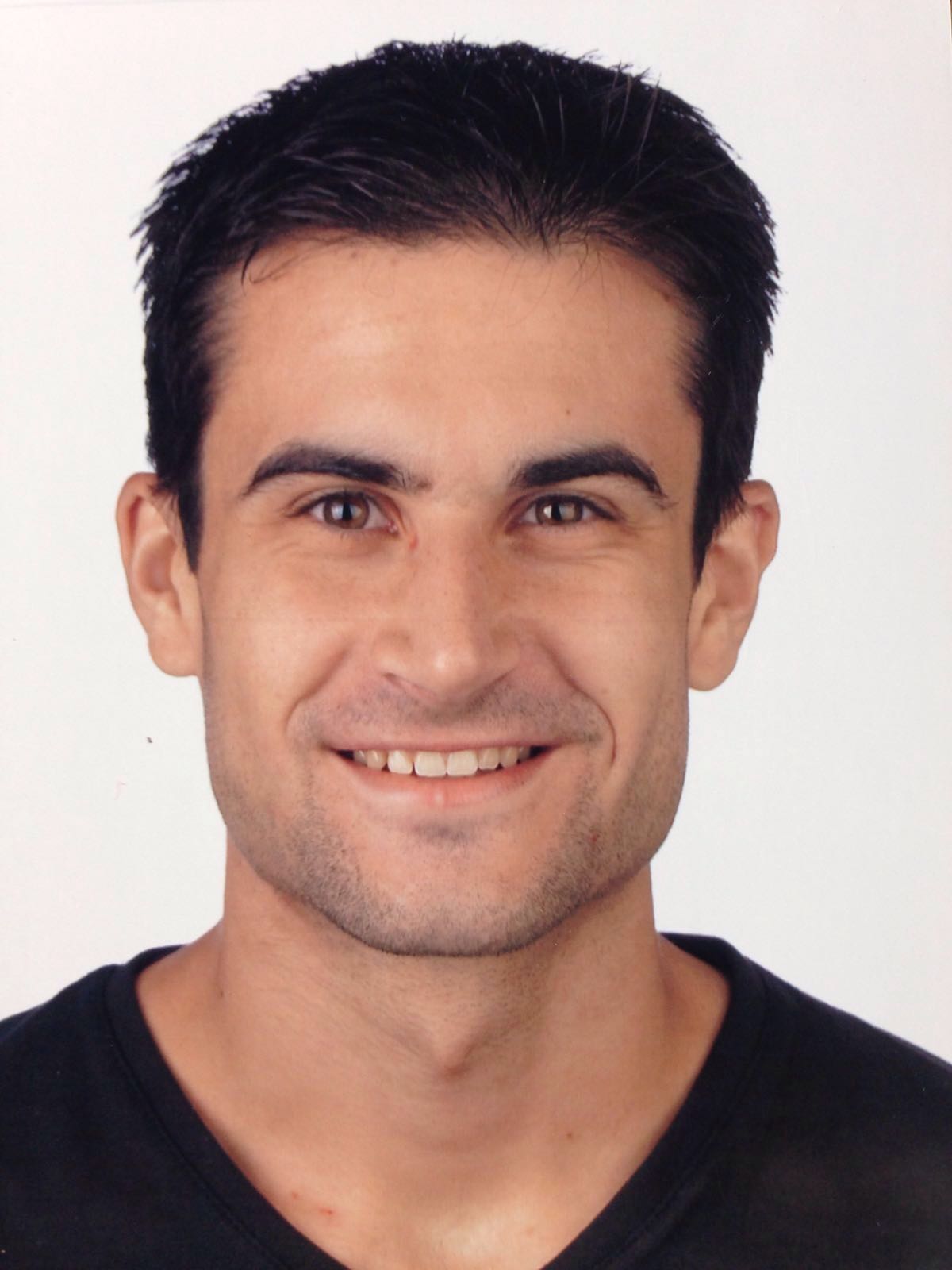}}

\begin{document}
\vspace*{4cm}
\title{Running vacuum versus the $\Lambda$CDM}

\author{Adri\`a G\'omez-Valent \footnote{E-mail: adriagova@ecm.ub.edu, sola@ecm.ub.edu, decruz@ecm.ub.edu . \label{Email}}\textsuperscript{,} \footnote{Speaker.}, Joan Sol\`a \textsuperscript{\ref{Email}}, and Javier de Cruz P\'erez \textsuperscript{\ref{Email}}}

\address{Departament de F\'isica Qu\`antica i Astrof\'isica, Univ. de Barcelona, Av. Diagonal 647, E-08028 Barcelona, Catalonia, Spain}

\maketitle\abstracts{
It is well-known that a constant $\Lambda$-term is a traditional building block of the concordance $\Lambda$CDM model. We show that this assumption is not necessarily the optimal one from the phenomenological point of view. The class of running vacuum models, with a possible running of the gravitational coupling G, are capable to fit the overall cosmological data
SNIa+BAO+H(z)+LSS+BBN+CMB better than the $\Lambda$CDM, namely at a level of $\sim 3\sigma$ and with Akaike and Bayesian information criteria supporting a strong level of statistical evidence on this fact. Here we report on the results of such analysis.}

\section{A brief introduction to Running Vacuum Models (RVM's)}

Some theoretical works point out that the dynamical nature of the vacuum could be described by QFT in curved space-time (see the lengthy and comprehensive review \cite{Sola13} and references therein).  In such proposal, the following ansatz for the $\beta$-function of $\Lambda$ is used:
\be
\frac{d\,\Lambda}{d\ln\mu}=\frac{1}{16\pi^2}\left[\sum_{i}C_i\mu^2+D_i\frac{\mu^4}{M_i^2}+...\right]\rightarrow \Lambda(H)=C_0+C_H H^2+C_{\dot{H}}\dot{H}+...
\ee
where in the second formula we identify the energy scale $\mu^2$ with a linear combination of $H^2$ and $\dot{H}$. Notice that there is no conflict at all with the Cosmological Principle, provided $\rho_\Lambda=\rho_\Lambda(H)$ is homogeneous and fully respects the general covariance. By virtue of the Bianchi identity, the covariant conservation of the full momentum tensor must be satisfied,
\be
\nabla^{\mu}(GT_{\mu 0})=0\rightarrow \frac{d}{dt}\left[G(\rho_m+\rho_r+\rho_\Lambda)\right]+3GH\sum_{i=m,r}(\rho_i+p_i)=0\,.
\ee
Thus, $\Lambda$ can vary if: (i) there exists an energy exchange between the vacuum and matter sectors: A-type models \cite{Atype,SoGoCr16}; (ii) the gravitational coupling G is also dynamical and matter is covariantly conserved: G-type models \cite{SoGoCr16,SoGoCr15}; (iii) (i)+(ii). Each of these scenarios gives rise to different background, linear and non-linear dynamics.
\begin{table*}
 \caption{Best-fit values for the $\Lambda$CDM and RVM's. 663 data points: SNIa+BAO+H(z)+LSS+BBN+CMB.}
\begin{center}
\resizebox{1\textwidth}{!}{
\begin{tabular}{| c | c |c | c | c | c | c| c | c |}
\multicolumn{1}{c}{Model} &  \multicolumn{1}{c}{$h$} &  \multicolumn{1}{c}{$\omega_b$} & \multicolumn{1}{c}{{\small$n_s$}}  &  \multicolumn{1}{c}{$\Omega_m$}&  \multicolumn{1}{c}{{\small$\nu_{eff}=\nu-\alpha$}}  &
\multicolumn{1}{c}{$\chi^2/dof$} & \multicolumn{1}{c}{$\Delta{\rm AIC}$} & \multicolumn{1}{c}{$\Delta{\rm BIC}$}
\\\hline {\small $\Lambda$CDM} & {\small $0.689\pm 0.004$} & {\small $0.02248\pm 0.00013$} & {\small$0.971\pm 0.004$} & {\small$0.300\pm 0.005$} & - & {\small $639.49/659$}  & - & -
\\\hline
{\small G1}  & {\small $0.676\pm 0.006$} & {\small $0.02243\pm 0.00014$} & {\small$0.969\pm 0.004$} & {\small$0.298\pm 0.004$} & {\small$0.0009\pm 0.0003$}  & {\small $622.14/658$} & {\small$15.35$} & {\small$10.85$}
\\\hline
{\small G2}  & {\small $0.676\pm 0.006$}& {\small $0.02242\pm 0.00014$} & {\small$0.968\pm 0.004$} & {\small$0.298\pm 0.004$} & {\small$0.0011\pm 0.0003$}  & {\small $621.85/658$} & {\small$15.64$} &  {\small$11.13$}
\\\hline
{\small A1}  & {\small $0.676\pm 0.006$}& {\small $0.02244\pm 0.00013$} & {\small$0.969\pm 0.004$} & {\small$0.298\pm 0.004$} & {\small$0.0009\pm 0.0003$}  & {\small $621.95/658$} & {\small$15.54$} & {\small$11.03$}
\\\hline {\small A2}& {\small $0.680\pm 0.005$}& {\small $0.02240\pm 0.00014$} & {\small$0.968\pm 0.004$} & {\small$0.299\pm 0.005$} & {\small$0.0012\pm 0.0004$}  & {\small $622.76/658$} & {\small$14.73$} & {\small$10.23$}
\\\hline
 \end{tabular}}
\end{center}
\label{tableFit1}
\end{table*}
One of the most striking features of these kind of models is that they are able to connect the primordial inflationary and the current accelerated phases of the Universe \cite{InflationSoGo}. After the inflationary epoch, the vacuum energy in the current Universe can be effectively described by $\rho_\Lambda(H,\dot{H})=\frac{3}{8\pi G}\left(c_0+\nu H^2+\frac{2}{3}\alpha\dot{H}\right)$, since the higher derivative contributions become totally negligible. The values of the two dimensionless parameters $\nu$ and $\alpha$ govern the running of the vacuum energy density and they are theoretically expected to be in the range $10^{-6}-10^{-3}$, depending on the extensions of the SM of Particle Physics and the predicted particle multiplicities in each of these extensions. Plugging $\rho_\Lambda$ in the Friedmann and the pressure equations, we can obtain the main background cosmological functions  \cite{Atype,SoGoCr16,SoGoCr15}.

\section{Akaike and Bayesian information criteria }

AIC and BIC are model selection criteria \cite{Stat}. They read, AIC$=\chi^2_{min}+2n$ and BIC=$\chi^2_{min}+n\,\ln N$, where $\chi^2_{min}$ is the minimum of the $\chi^2$ function, $N$ is the number of data points and $n$ the number of free parameters that enter the fit. AIC and BIC allow us to compare models with a different number of free parameters that have been fitted with the same data set, strongly penalizing the use of extra $dof$. To test the effectiveness of a given RVM (versus the $\Lambda$CDM) for describing the overall data, we evaluate the pairwise differences $\Delta$AIC ($\Delta$BIC) with respect to the model that carries smaller value of AIC (BIC), in this case the RVM. The larger these differences the higher is the evidence against the model with larger value of $\Delta$AIC ($\Delta$BIC). For $\Delta$AIC and/or $\Delta$BIC above 10, one speaks of ``very strong evidence''. According to both criteria, the results of Table 1 show that the $\Lambda$CDM appears very strongly disfavored as compared to the RVM's. Hence, the RVM's emerge as serious alternative candidates for the description of our expanding Universe \cite{SoGoCr16,SoGoCr15}. The dynamical character of the vacuum energy density in them seems to be firmly supported by the current cosmological data, in which the absence of vacuum dynamics is excluded at $\sim 3\sigma$.

\section*{Acknowledgments}

A. G\'omez-Valent acknowledges the Quantum Physics and Astrophysics Department of the Univ. of Barcelona, which has made possible his attendance to the $51st$ Rencontres de Moriond.

\end{document}